\overfullrule=0pt
\def\pz{\phantom{0}}

\input stromlo

\title The Kinematics of Dwarf Spheroidal Galaxies

\shorttitle Dwarf Spheroidal Galaxies

\author Mario Mateo

\shortauthor Mateo

\affil University of Michigan, Ann Arbor, MI 48109, USA

\abstract I review observational data on the kinematic properties of
dwarf spheroidal galaxies in the halo of the Milky Way and beyond.  The
present data confirm previous claims that these small galaxies have 
unusually large central velocity dispersions. `Simple'
sources of bias such as binary stars, internal
atmospheric motions, measurement errors and small sample sizes cannot 
explain the large dispersions measured in all dSph systems.
Recent data suggest that
in some of these dwarfs the velocity dispersion profiles are flat out to their
classical tidal radii.  I discuss how these results can (or cannot) be 
understood by invoking a variety of distinct models, including classical dark 
matter halos, tidal disruption, and MOND.

\section Introduction

Dwarf spheroidal (dSph)  
galaxies have provided a host of surprises ever since 
their `renaissance' in the early 1980's.   Led early on by Marc Aaronson,
it has become increasingly obvious
that these fragile-looking galaxies of the outer
halo appear to have complex star-formation histories (Mould and Aaronson
1983; see Gallagher and Wyse 1994 for a review), may contribute significantly
to the stellar population of the outer halo (Kunkel 1979; Lynden-Bell 1982;
Ibata et al. 1994; see a review by Mateo 1996), and may shed fundamental
insights on the nature of dark matter  in galaxies.  
My principal goal here is to focus on the latter topic and to
provide an update on the kinematic 
studies of dSph galaxies.   I'll restrict myself to studies of resolved dSph 
systems, which at present means only galaxies in the halo of our Galaxy and 
perhaps just out to the dwarf companions of M~31.  

\section Kinematics of Dwarf Spheroidal Galaxies

I will start this summary where an earlier review (Mateo 1994)
left off.  Please see that paper for a summary of the observational
data on dSph galaxies through about 1993 (see also Gallagher and
Wyse 1994).  The basic nature of kinematic studies of these galaxies has
begun to change qualitatively since my last review.  Now, multi-object and
near-IR spectroscopy has made more of an impact, and serious efforts are
underway to extend observations to the very outer limits of individual 
galaxies and to ever more distant galaxies using new 8-10m class
telescopes.  Because it is often useful to have the kinematic and
structural data in one place,
I provide these in Table~1 which is taken from Table~1 of Mateo (1994).
I've also included results for systems that
have either been discovered (Sgr; Ibata et al. 1994; Mateo et al. 1995; 
Ibata et al. 1997) or first studied kinematically in the
optical (LGS~3; Lo et al. 1993; Lee 1995) since that earlier review.
Irwin and Hatzidimitriou (1995) provide a recent compilation of dSph 
structural parameters; their results are consistent with the data listed
in Table~1.
Table~2 summarizes the kinematic results from a number of relatively 
recent studies of resolved dSph galaxies; earlier references can be found
in Mateo (1994).

\table 1. Properties of Local Group dSph Galaxies

@l @c @c @c @c @c @c \\
\tworules
&\c{Name} & \c{$D$} & \c{$R_{GC}$} & \c{$M_V$} & \c{$R_c$} & \c{$\Sigma_{0,V}$}
  & \c{$I_0$} \\
&  & \c{kpc} & \c{kpc}  &  \c{mag} &  \c{pc}   & \c{mag arcsec$^{-2}$} & 
  \c{\L$_\odot$ pc$^{-3}$} \\
\onerule
& Draco & \pz76 & \pz76 & \pz$-8.7$ & 190 & 25.2 & 0.016 \\
& Carina & \pz87 & \pz89 & \pz$-8.9$ & 210 & 25.2 & 0.011 \\
& Ursa Minor & \pz63 & \pz66 & \pz$-8.9$ & 290 & 25.1 & 0.009 \\
& Sextans & \pz88 & \pz91 & $-10.0$ & 380 & 25.5 & 0.004 \\
& And~III & 690 & \pz59 & $-10.2$ & 250 & 24.5 & 0.011 \\
& Leo~II  & 220 & 220     & $-10.2$ & 220 & 23.8 & 0.037 \\
& LGS~3 & 810 & 170 & $-10.4$ & 190 & 24.8 & 0.011 \\
& Sculptor & \pz78 & \pz78 & $-10.7$ & 200 & 24.1 & 0.023 \\
& And~I & 690 & \pz39 & $-11.7$ & 315 & 24.4 & 0.009 \\
& And~II & 690 & 123 & $-11.7$ & 330 & 24.5 & 0.008 \\
& Leo~I   & 230 & 230     & $-11.7$ & 150 & 22.3 & 0.13\pz \\
& Sagittarius & \pz25 & \pz16 & $-13.0$ & 550 & 25.3 & 0.001 \\
& Fornax & 131 & 133 & $-13.7$ & 640 & 23.2 & 0.020 \\
& NGC~147 & 690 & \pz90 & $-15.1$ & 230 & 21.6 & 0.18\pz \\
& NGC~185 & 690 & \pz86 & $-15.3$ & 200 & 20.9 & 0.37\pz \\
& NGC~205 & 690 & \pz\pz8 & $-16.3$ & 170 & 19.9 & 1.2\pz\pz \\
\onerule
\space\space
\annot $D$ = heliocentric distance. \\
\annot $R_{GC}$ = Galactocentric distance or projected distance for M31
satellites. \\
\annot $M_V$ = absolute visual magnitude. \\
\annot $R_c$ = the approximate King core radius. \\
\annot $\Sigma_{0,V}$ = central V surface brightness. \\
\annot $I_0$ = central luminosity density. \\

\table 2. Recent Kinematic Studies of dSph Galaxies

@l @c @c @c @c @l \\
&\c{Name} & $N_*$ & $N_{obs}$ & $\sigma_0$ & $M/L_V$ & \c{Reference} \\
& & & & km/s & & \\
\onerule
& Carina & 17 & 0 & $6.8 \pm 1.6$ & $39 \pm 23$ & see M94 \\
& Draco & 91 & 167 & $8.5 \pm 0.7$ & $57 \pm 9$ & AOP95 \\
& & 17 & 17 & $10.5 \pm 2.0$ & $166 \pm 100$ & H96 \\
& Fornax & 37 & 45 & $9.9 \pm 1.7$ & $12 \pm 5$ & see M94 \\
& & 215 & 250 & $9.6 \pm 1,8$ & $11 \pm 5$ & in preparation\\
& Leo~I & 34 & 40 & 9.0 $\pm$ 1.2 & $8 \pm 3$ & in preparation \\
& Leo~II  & 31 & 37 & $6.7 \pm 1.1$ & $12 \pm 3$ & V95 \\
& LGS~3 & 4 & 4 & $7.0 \pm 4.0 $ & $20 \pm 15$ & in preparation \\
& Sagittarius & $\sim 300$ & $\sim 450$ & $11.4 \pm 0.7$ & $\sim 100$ & I97 \\
& Sculptor & 32 & 32 & $7.0 \pm 1.2$ & $9 \pm 4$ & see M94 \\
&  & 23 & 23 & $6.2 \pm 1.1$ & $9 \pm 6$ & QDP95 \\
& Sextans & 33 & 70 & $6.2 \pm 0.8$ & $18 \pm 10$ & S93 \\
& & 21 & 30 & $7.0 \pm 1.2$ & $124 \pm 70$ & H94a \\
& Ursa Minor & 94 & 206 & $8.8 \pm 0.8$ & $55 \pm 10$ & AOP95 \\
& & 35 & 44 & $6.7 \pm 1.0$ & $59 \pm 30$ & H94b \\
\onerule
\space
\annot $N_*$ = number of different stars observed. \\
\annot $N_{obs}$ = number of total observations. \\
\annot $\sigma_0$ = central velocity dispersion. \\ 
\annot $M/L_V$ = central V-band mass-to-light ratio in solar units. \\
\annot Abbreviations used here are included with the formal references. \\

Some of the highlights of these most 
recent kinematic studies for individual dSph galaxies are listed below:

\vskip0.5em

\noindent {\bf Fornax}  -- 
Mateo and Olszewski (1997, in preparation) 
have obtained 
precise velocities of giants located
out to and even somewhat beyond the formal tidal radius of Fornax.  These
data demonstrate that (a) Fornax is possibly losing some stars to the
surrounding halo field, and (b) the velocity dispersion profile of the 
galaxy is remarkably flat, inconsistent with the profile expected for
a King model that best fits the surface-brightness profile of the galaxy
(see Figure~1).   These results are based on observations of 215
Fornax members.

\noindent {\bf Sculptor}  -- Queloz et al. (1995) published a study of 23 
K giants in this southern dSph based on Echelle spectra at ESO.  They 
reobserved many of the stars studied by Armandroff and Da~Costa (1986), and
found two of these to be significant velocity variables.  With the probable
binaries removed, the two datasets agreed well.  Da Costa and Armandroff
(private communication) see no significant change in the velocity
dispersion at two radial positions in Sculptor.

\noindent {\bf Ursa Minor}  -- Two groups have published excellent 
new kinematic results for both this galaxy and Draco.  Olszewski et al. (1995,
1996) and Armandroff et al. (1995) have obtained the largest samples of stars
yet published for an dSph galaxies; for UMi, 206 observations were obtained
of 94 member stars using a variety of instruments and telescopes.  Within
their datasets, no significant systematic errors were observed.  However,
these authors do find differences with the results of Hargreaves
et al. (1994b), but only at the 1-2$\sigma$ level in the velocity dispersion
and the rotation.  Both studies find significant evidence for rotation in
UMi, but oddly, more nearly along the {\it minor} axis.  The rotation
amplitude is much less than the central velocity dispersion, with 
$v_{rot}/\sigma_0 \sim 0.3$.  The study by Olszewski et al. (1995) is 
particularly noteworthy because it reports eight years of 
high-precision velocity measurements of UMi and Dra stars.  Velocity variables
are certainly present; in all, four good candidates are identified in
UMi as probable binaries.  But the long-term effects of atmospheric
velocity variability appears to be negligible for {\it bona fide} K 
giants in UMi.

\noindent {\bf Draco}  -- Olszewski et al. (1996) identify two velocity
variables that may be binaries.  The sample size in Draco, as for UMi, was
greatly increased from earlier work to 167 observations of 91 member stars.
Unlike UMi, there is no apparent rotation seen in the Draco kinematic data.
Hargreaves, et al. (1996b) have also recently studied Draco, reaching
statistically similar results.

\figureps[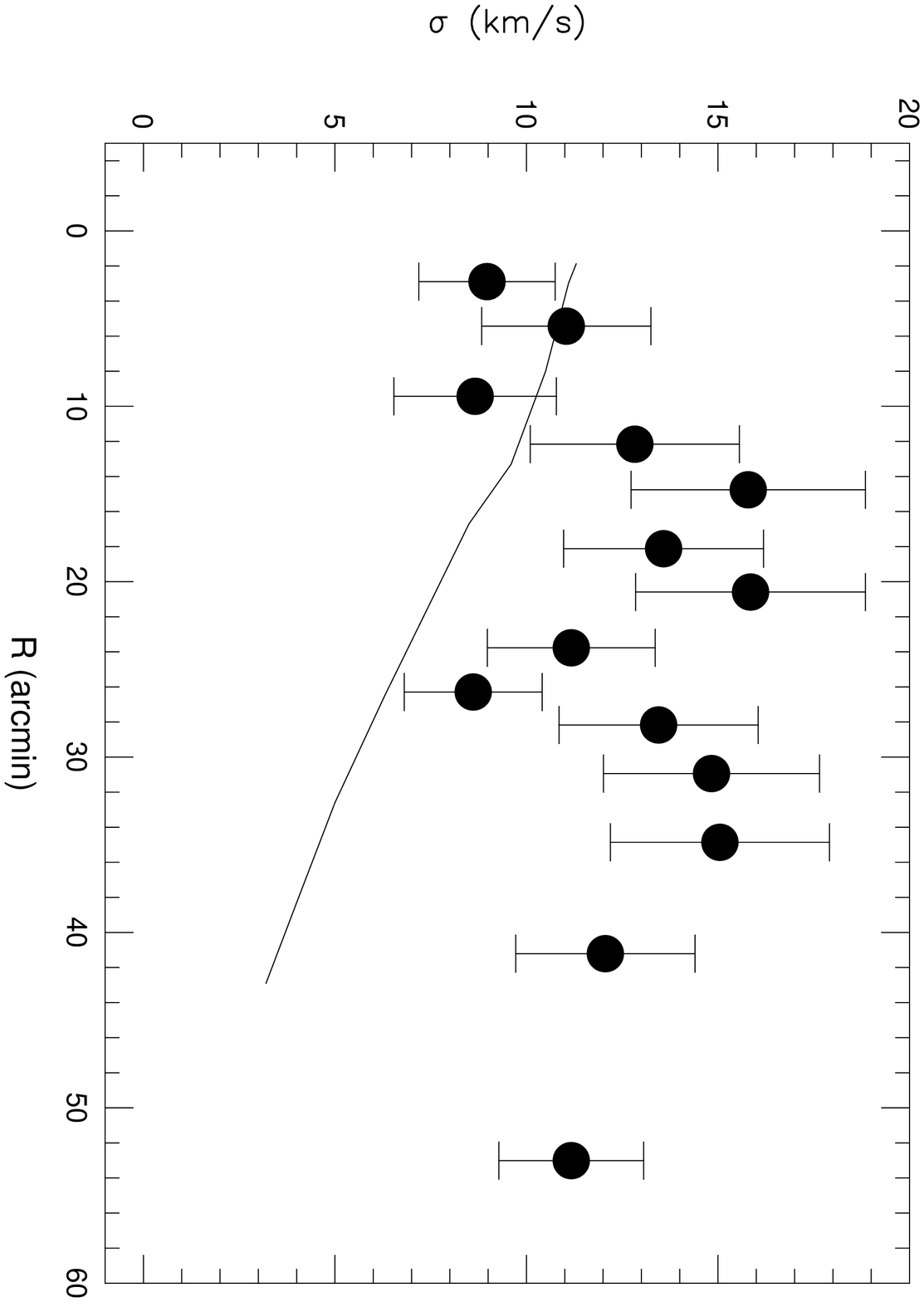,0.6\hsize] 1. The velocity dispersion profile
of Fornax.  Each bin is based on measurements of 15 galaxy members, 
except for the outer bin which contains 20 stars.

\noindent {\bf Leo~II}  -- Vogt et al. (1995) observed Leo~II using the
HIRES spectrograph on the Keck 10m telescope and obtained precise velocities
of 31 red giants, along with repeat measurements to assess the velocity
precision.   Because it is located in the outer halo,
tidal effects should not affect the kinematics of Leo~II, and it
was chosen as a test case to distinguish between a dark-matter  or tidal
origin for the central velocity dispersion.   In the end, the derived
M/L was neither unusually high, nor low, but rather `just right';
see section 4 and Figure~2.

\noindent {\bf Leo~I}  -- The same group that studied Leo~II obtained Keck
data for 34 red giants in Leo~I; again, the velocity precision is high and
was explicitly checked with repeat measurements of Leo~I giants.  We 
confirm the large systemic velocity of Leo~I (the implications of which are
described by Zaritsky et al. 1989) and derived the dispersion listed in 
Table~2.  As for Leo~II, Leo~I is sufficiently far out in the halo 
that tidal effects cannot plausibly have affected its internal kinematics.
However, Leo~I's higher luminosity means that its central velocity dispersion
was not expected to be very small even with no dark matter. 

\noindent {\bf Sagittarius}  -- Discovered from its kinematic signature
within a survey of the Galactic Bulge (Ibata et al. 1994),
Sgr provides a unique test case of how a small galaxy dissolves into the 
halo of its larger parent.  Because of this, the analysis of Sgr is also
qualitatively different from that of all the other dwarfs.  In particular,
one cannot expect to use equilibrium models to properly represent the
velocities of stars in Sgr, nor can one ignore the very large spatial 
extent of the galaxy (Mateo et al. 1996; Alard 1996; 
Fahlman et al. 1996) and the purely geometric
effects on the systemic velocities 
due to this large angular extent.  The results listed in Table~2 -- taken
from Ibata et al. 1997 -- 
must technically be qualified by the locations at which the measurements were
made.  However, these authors find relatively little 
change in the dispersion as a function of distance along the major axis of
Sgr along about 10$^\circ$ of its major axis.  
As this and previous studies have clearly
demonstrated (Allen and Richstone 1988;  Moore and Davis 1994; 
Piatek and Pryor 1995; Oh et al. 1995;
Johnston et al. 1995; Velazquez and White 1995), 
the very large extent of Sgr means that detailed
models of the interaction of this dwarf 
with the Galaxy {\it must} be used to interpret the kinematic measurements.

\noindent {\bf Sextans}  -- Hargreaves et al. (1994a) broadly confirm the 
observational results of Suntzeff et al. (1993), but because these authors
used different structural parameters and the stellar samples are small, the
resulting M/L ratios differ by an apparently large amount (however,
note that the errors are also large).  Hargreaves et al. (1994a) find no 
evidence for significant rotation in Sextans.

\noindent {\bf LGS~3}  -- K. Cook and C. Stubbs obtained Keck spectra
of four bright giants in this dwarf located near M31/M33  
 which 
E. Olszewski and I reduced to obtain the 
velocity dispersion listed in Table~2.  Because the sample is so small
(four stars) and the velocity precision is modest, we can only conclude 
that (a) the stellar velocity dispersion is not radically
different from that of the meager amount of H~I gas in LGS~3 (Lo et al. 1993),
and (b) we are at the dawn of an exciting era
when 8-10m class telescopes will allow
us to study stellar kinematics of dwarfs up to 1 Mpc distant!

\section Can We Trust These Results?

The closest dSph systems are about 70 kpc away and are composed of
principally old (age $\geqsim$ 1 Gyr) stars.   The 
central velocity dispersions of
these galaxies are expected to be only about 1-3 km/s {\it if} they contain
no dark matter, are unaffected by tidal effects, and if normal Newtonian
gravity applies to these very loose, low-surface brightness systems.
This all means that we are forced to try to measure precise ($\epsilon_{obs}
\leqsim 2$ km/s) velocities of {\it many} stars fainter than V $\sim$ 17.5,
and sometimes in the presence of significant field-star contamination.

Because of these difficulties
and because the implications of large central velocity
dispersions in dSph galaxies are potentially far-reaching, there
has historically -- and rightfully -- been concern about the
reliability of these results.   Aaronson's first paper helped set 
this sceptical tone: he
based his estimate of the
velocity dispersion of Draco (Aaronson 1983) on measurements of
three stars --  upgraded to four in a note added in proof!

It should be clear from Table~2 that sample sizes are now no longer a 
serious concern in kinematic studies of dSph systems.  The worst case 
(for a Galactic dwarf) is Carina.  Larger samples remain
important, however, to study the radial velocity dispersion profiles
in dwarfs, and to constrain the anisotropy of the velocity distribution
of stars in these galaxies.

A second concern has been the influence of binary stars.  An early review by
Aaronson and Olszewski (1985) described simulations that
suggested that binaries are unlikely to inflate 
the true dispersion to the large observed values in objects such as 
Draco and Ursa Minor.  This result was consistent with the meager 
direct information on the frequency of binaries in the period range 
relevant for red giants in dSph galaxies (e.g. 
Olszewski et al. 1995).  More recently,
Olszewski et al. (1995, 1996) and Hargreaves et al. (1996a)
completed comprehensive simulations
of the effects of binaries in dwarf galaxies.  Both studies compared the
observed vs. true dispersion for samples with and without binaries over
large period ranges.  Both concluded that the observed binary fractions and
a reasonable period distribution cannot account for the dispersions measured 
in any dSph galaxy.  The dispersions have {\it not} been inflated by
orbital motions in undetected binary systems.

Another possible systematic source of bias in the dispersions is
atmospheric motions in luminous giants.  This so-called `jitter' is not
severe in first-ascent giants in older systems such as dSph galaxies, and
it has long been appreciated that the fainter giants are less 
susceptible to these velocity excursions (Pryor et al. 1988). In practice, most
observers have tried hard to obtain data only of the red giants as far
below the red giant branch tip as possible.
Stars above this luminosity, such as asymptotic giant
branch (AGB) C stars, are less reliable, but they remain useful as the 
most luminous stars in these galaxies, and thus the easiest to observe 
in the most distant systems.  Olszewski et al. (1995) show empirically
that the long-term effects of atmospheric motions in dSph giants are 
negligible.

This summary should make it clear that modern estimates of the velocity
dispersions in dSph galaxies {\it are} generally reliable at the 1-2 km/s 
precision level.  There are two immediate conclusions to draw from
this.  First, {\it no} dSph galaxy with adequate observations (see Table~2)
has a central velocity dispersion smaller than 6 km/s. Second, the
current measurements cleanly rule out velocity dispersions as low as
one would expect by merely 
scaling globular cluster kinematical results by the structural
parameters of dSph galaxies (Richstone and Tremaine 1988).  There is a
fundamental phenomenon reflected in the kinematics of dSph galaxies that
we must understand.

\section The Origin of dSph Dispersions

This of course begs the question:
What is the nature of this fundamental kinematic difference between dwarfs and
globulars -- dark matter content, tides, anisotropy, the nature of gravity?
Or something else?  I wish I could answer this, and the frustration with
each of these explanations at different times causes me to personally
waver on 6-month timescales.  Let's consider the possibilities in order
of increasing speculativeness (by today's reckoning):

\vskip0.5em

\noindent {\it Dark Matter}  -- 
The `standard' physical interpretation for the large
dispersions in dSph galaxies is that they contain large quantities of dark 
matter (DM).  The recent observations in Fornax, Sagittarius and to a lesser
extent, Ursa Minor, Draco and Sculptor, suggest that if DM is the culprit,
it is distributed in a more extended manner than the visible stars.
Ibata et al. (1997) in particular argue that the 
DM in Sagittarius must be particularly extensive to have stabilized that
galaxy for many close passages with the Milky Way.  Thus, DM halos are 
large.  Pryor and Kormendy (1990) noted that
the inferred DM central densities are amazingly
high, perhaps as large as 1 M$_\odot$/pc$^3$ in systems with luminous
matter densities of only 1-5\% of this value.

Mateo et al. (1993) pointed out that all dSph systems 
had inferred DM halos of remarkably similar mass, about
1-5 $\times 10^7$M$_\odot$.  Figure~2 illustrates this feature along with 
new
data from additional systems.  The dashed line represents the
equation $M_{tot,0} = 5.0L_{tot} + 2.5 \times 10^7$M$_\odot$, where $M_{tot}$
is the `total' mass, $L_{tot}$
is the total luminosity, and the factor `5.0' is taken as the standard
V-band M/L ratio of the visible stars in the galaxy -- all quantities are in
solar units.  In all cases
(except Sagittarius), the 
masses assume that mass follows light and isotropic
orbits (Richstone and Tremaine 1988), so they represent the
total masses of the galaxies as derived from their central
kinematic/structural properties.  Nonetheless, the in Figure~2
correlation is striking, perhaps telling us that DM halos in dSph systems
are somehow intimately related and, because of the deviant location of
Sagittarius, that severe tidal
effects do ultimately affect the inferred M/L ratios (see also Bellazzini
et al. 1996).  
Do such minimum-sized
`chunks' of DM that contribute to the halos of larger galaxies, groups and
clusters?

\figureps[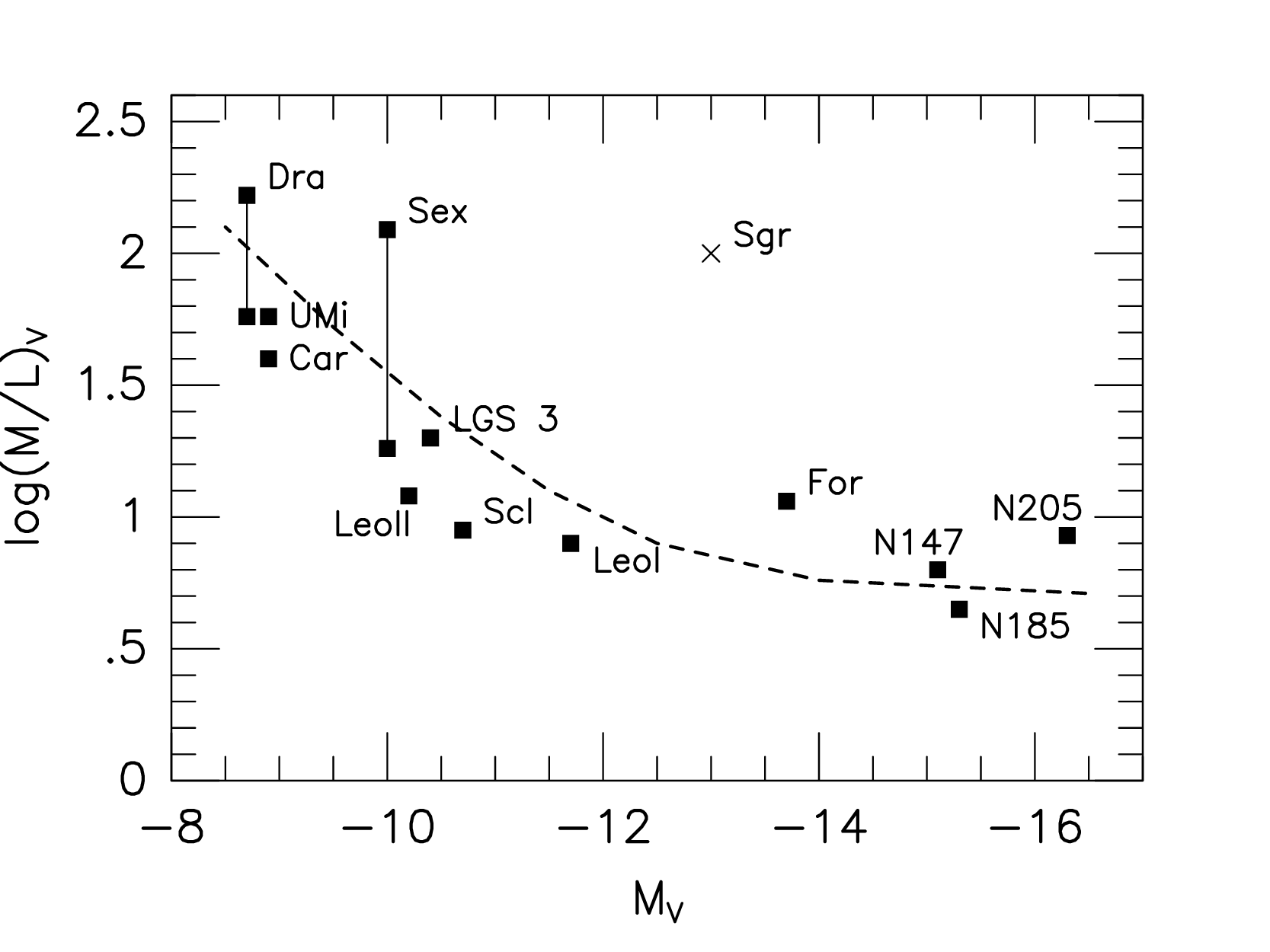,0.7\hsize] 2. The correlation between M/L$_V$ and $M_V$ for
Local Group dSph galaxies with good kinematic data.  The dashed line is
equivalent to a model where each galaxy has a dark halo mass of
$2.5 \times 10^7$M$_\odot$ plus a luminous component with M/L$_V$ = 5.0.
Sagittarius is noted with a cross.  The data for the Andromeda dwarfs 
(apart from LGS~3) come from the references in Mateo 1994.

It is interesting to note that dSph galaxies provide one of the
only constraints on what DM isn't.  As first noted by Faber and Lin (1983)
and updated by Gerhard and Spergel (1992),
phase space restrictions demand that if neutrinos provide the DM in these
galaxies, the neutrino mass must be much larger than current experimental
upper limits. 

\vskip0.5em

\noindent {\it Galactic Tides}  -- Hodge and Michie (1969) noted that
the structural parameters of Ursa Minor indicate that it may be
in the terminal phases of complete disruption due to severe tidal interaction
with the Milky Way.  
That some dSph galaxies are affected by Galactic tides is
indisputable.  Sagittarius is a clear example, but Carina (Kuhn et al. 1996),
and possibly Sextans (Gould et al. 1992) have had member stars detected 
far from the main bodies of the galaxies.  Radial surface
density profiles 
of dSph systems (Irwin and Hatzidimitriou 1995) suggest that many of these
galaxies seem to `melt' into the surrounding halo.

Nonetheless, 
for many people, kinematic measurements `confirmed' suspicions that
DM stabilizes these galaxies against disruption.  Recent
models (particularly Piatek and Pryor 1995 and Oh et al. 1995) argue that
tides could not possibly inflate the {\it central} dispersions to the observed
values without also observing strong streaming motions (that would be 
interpreted as rotation).  Pryor (1996) offers a recent review of these
studies and their conclusions.

Another class of models directly challenges the conclusion that
dSph systems are DM dominated.  Kuhn and
Miller (1989) and Kroupa (1996) argue that resonances between the internal
oscillation periods of dSph galaxies induced by tides and their orbital
periods could make systems with large central dispersions persist for extended
periods.  It would be of great interest to widely confirm and expand on these
models to understand if the resulting dwarfs are indeed sufficiently 
long-lived to be seen as (mostly) non-rotating dwarfs with large central
velocity dispersions.  It is hard to understand how tides account for
Figure~2, though small-number statistics could conceivably still be important.

\vskip0.5em 

\noindent {\it Modified Gravity}  --  This is clearly the currently 
least popular
option at present. Milgrom has argued
that MOND (MOdified Newtonian Gravity) offers an alternative explanation of
the observed kinematical results for dSph galaxies if the observational
uncertainties are realistically accounted for 
(Milgrom 1995; see Gerhard 1994 for a different view).  Even one 
undisputed failure of MOND would disqualify it, unless we also wish to 
consider a non-universal modified gravity law!  But 
disagreements remain whether such failures are observed, while some impressive
successes have been noted among large high- and low-surface brightness
galaxies (McGaugh, private communication; Mannheim 1997).  
My point here is merely that
dSph galaxies are not obviously inconsistent with Milgrom's version of
MOND, a concept
originally designed to explain rotation curves of galaxies thousands
of times more luminous than dSph systems.

\references

Aaronson, M. 1983, ApJ, 266, L11

Aaronson, M., \& Olszewski, E. W. 1985, IAU Symp 153, p. 159

Alard, C. 1996, ApJ, 458, L17

Allen, A. J., \& Richstone, D. O. 1988, ApJ, 325, 583

Armandroff, T. E., \& Da Costa, G. S. 1986, AJ, 92, 777

Armandroff, T. E., Olszewski, E. W., \& Pryor, C. 1995, AJ,
110, 2131 (AOP95)

Bellazzini, M., Fusi Pecci, F.,
Ferraro, F. R. 1996, MN, 
278, 947

Faber, S. M., and Lin, D. N. C.
1983, ApJ, 266, L21

Fahlman, G. G., Mandushev, G., Richer, H. B., Thompson, I. B.,
\& Sivaramakrishnan, A. 1996, ApJ, 459, L65

Gallagher, J. S., \& Wyse, R. F. G. 1994, PASP, 106, 1225

Gerhard, O. E. 1994, 1994, ESO/OHP Workshop, G. Meylan and P. Prugniel, eds.,
p. 335

Gerhard, O. E., \& Spergel, D. N.
1992, ApJ, 389, L9

Gould, A., Guhathakurta, P., Richstone, D., \& Flynn, C. 1992,
ApJ, 388, 345

Hargreaves, J. C., Gilmore, G., and Annan, J. D. 1996a, MN, 279,
108

Hargreaves, J. C., Gilmore, G., Irwin, M. J., \& Carter, D. 1994a, MN, 269, 957
(H94a)

Hargreaves, J. C., Gilmore, G., Irwin, M. J., \& Carter, D. 1994b, MN, 271, 693
(H94b)

Hargreaves, J. C., Gilmore, G., Irwin, M. J., \& Carter, D. 
1996a, MN, 282, 305 (H96)

Hodge, P. W., \& Michie, R. W. 1969, AJ, 74, 587

Johnston, K. V., Spergel, D. N., \& Hernquist, L. 1995, ApJ, 
451, 598

Kroupa, P. 1996, preprint [astro-ph/9612028]

Kuhn, J. R., \& Miller, R. H. 1989, ApJ, 339, L41

Kuhn, J. R., Smith, H. A., \& Hawley, S. L. 1996, 
ApJ, 469, L93

Ibata, R. A., Gilmore, G., \& Irwin, M. J. 1994, Nature, 370, 194

Ibata, R. A., Wyse, R. F. G., Gilmore, G., Irwin, M. J., \& Suntzeff, N. B.
1997, AJ, in press [astro-ph/9612025] (I97)

Irwin, M., \& Hatzidimitriou, D. 1995, MN, 277, 1354

Kunkel, W. E. 1979, ApJ, 228, 718

Lee, M.-G. 1995, AJ, 110, 1129

Lo, K. Y., Sargent, W. L. W., \& Young, K. 1993, AJ, 106, 
507

Lynden-Bell, D. 1982, Observatory, 102, 202

Mannheim, P. D. 1997, 18th Texas Symposium

Mateo, M. 1994, ESO/OHP Workshop, G. Meylan and P. Prugniel, eds.,
p. 309 (M94)

Mateo, M. 1996, ASP Conf. 92, H. Morrison and A. Sarajedini,
eds., p. 434

Mateo, M., Mirabal, N., Udalski, A., Szymanski, M., Kaluzny,
J., Kubiak, M., Krzeminski, W., \& Stanek, K. Z. 1996, 
ApJ, 458, L13

Mateo, M., Olszewski, E. W., Pryor, C., Welch, D. L., \& Fischer,
P. 1993, AJ, 105, 510

Milgrom, M. 1995, ApJ, 455, 439

Moore, B., \& Davis, M. 1994, MN, 270, 209

Mould, J., \& Aaronson, M. 1983, ApJ, 273, 530

Oh, K. S., Lin, D. N. C., \& Aarseth, S. J. 1995, ApJ,
442, 142

Olszewski, E. W., Aaronson, M., \& Hill, J. M. 1995, 110, 2120

Olszewski, E. W., Pryor, C., \& Armandroff, T. E. 1996, 
AJ, 111, 750

Piatek, S., \& Pryor, C. 1995, AJ, 109, 1071

Pryor, C. 1996, ASP Conf. 92, H. Morrison and A. Sarajedini,
eds., p. 424

Pryor, C., \& Kormendy, J. 1990, AJ, 100, 127

Pryor, C., Latham, D. W., \& Hazen, M. L. 1988, AJ, 96, 123

Queloz, D., Dubath, P., \& Pasquini, L. 1995, A\& A, 300, 31 (QDP95)

Richstone, D. O., \& Tremaine, S.
1988, ApJ, 327, 82

Schweitzer, A. E., \& Cudworth, K. M. 1996, ASP 
Conf. 92, H. Morrison and A. Sarajedini, eds., p. 532

Schweitzer, A. E., Cudworth, K. M., Majewski, S. R., \&
Suntzeff, N. B. 1995, AJ, 110, 2747

Suntzeff, N. B., Mateo, M., Terndrup, D. M., Olszewski, E. W.,
Geisler, D. W., \& Weller, W. 1993, ApJ, 418, 208 (S93)

Velazquez, H., \& White, S. D. M. 1995, MN, 275, L23

Vogt, S., Mateo, M., Olszewski, E. W., \& Keane, M. J. 1995,
AJ, 109, 151 (V95)

Zaritsky, D., Olszewski, E. W., Schommer, R. A., Peterson, R.
C., \& Aaronson, M. 1989, ApJ, 345, 759

\bye